\documentclass{neuthist18}

\def\be{\begin{equation}}
\def\ee{\end{equation}}
\def\bea{\begin{eqnarray}}
\def\eea{\end{eqnarray}}

\newcommand{\Photo}{}

\begin{document}
\vspace*{4cm}
\title{REACTOR NEUTRINOS: TOWARD OSCILLATIONS \\ 
\bigskip
Invited Talk at the "History of the Neutrino" Conference, September 2018, Paris }

\author{ Petr Vogel }

\address{Kellogg Radiation Laboratory and Department of Physics 256-48\\
1200 E. California Bl., Pasadena CA 91125, USA}

\maketitle\abstracts{
I shall sketch the history of reactor neutrino physics over
five decades since the Reines-Cowan proof of neutrino
existence in the late 50s, till the advent of the present
era of precision reactor neutrino oscillation experiments.
There are three chapters of this story:
i) Exploration of possibilities of the reactor neutrinos in the 60s and 70s;
ii) Looking for oscillations under the streetlamp in the 80s and 90s;
iii) Exploring oscillations in detail with known (almost) $\Delta m^2_{atm}$ and $\Delta m^2_{sol}$.}

\section{Introduction}

Nuclear reactors are powerful sources of low energy electron antineutrinos. The $\bar{\nu}_e$ are produced in the $\beta$ decay
of the neutron rich fission fragments. Each of the fragments produces on average 3 $\bar{\nu}_e$; there are thus $\sim$ 6 $\bar{\nu}_e$
per each act of fission. A typical power reactor of 3 GW thermal power undergoes $\sim 10^{20}$ fissions/s and thus produces 
$\sim 6\times 10^{20}$ $\bar{\nu}_e$/s with energies typical for nuclear $\beta$ decay.  The $\bar{\nu}_e$ spectrum fast decreases
with increasing energy, there are only very few $\bar{\nu}_e$ left past 8 MeV. 

Neutrinos interact with anything else only by weak interactions. It is easy to estimate the order of magnitude of the corresponding
cross section. At low energies ($\sim$ MeV) the cross section can depend only on the $\bar{\nu}_e$ energy, and must be proportional
to $G_F^2$ ($G_F = 1.17 \times 10^{-11}$ MeV$^{-2}$). By dimensional arguments then $\sigma \sim G_F^2 E^2 (\hbar c)^2  \sim 10^{-44}$
cm$^2$. This is, basically, the argument first made by Bethe and Peierls already in 1934 \cite{bethe34}.

Electron antineutrino at nuclear reactors can interact by both charged and neutral weak currents. The following
four reactions have been successfully observed: 
\begin{table}[h]
\begin{center}
\begin{tabular} {|c|c|c|l|}
\hline
reaction & label & $\sigma$ (10$^{-44}$cm$^2$/fission) & threshold (MeV) \\
\hline
$\bar{\nu}_e + p \rightarrow e^+ + n$  &   ccp  & 63 &    1.8  \\
$\bar{\nu}_e + d \rightarrow e^+ + n + n$ &  ccd &1.1 &    4.0  \\
$\bar{\nu}_e + d \rightarrow  \bar{\nu}_e + p + n $ & ncd &  3.1 & 2.2 \\
$\bar{\nu}_e + e^- \rightarrow \bar{\nu}_e + e^- $ & el.sc. & 0.4 & 1-6 (range) \\
\hline
\end{tabular}
\end{center}
\end{table}

Neutrino capture on protons (ccp), usually called `Inverse Beta Decay' (IBD), has obvious advantages. It has
much larger cross section and the neutron recoil energy is low  so that the positron carries essentially all available
energy. The neutron is eventually captured either on one of the target protons (2.2 MeV capture energy) or
is captured on some added admixture with very large neutron capture cross section, such as Gd (8.0 MeV cascade of $\gamma$).
The space and time coincidence requirement between the positron detection and neutron capture $\gamma$ rays is a powerful
tool for background suppression. The IBD reaction was first used in the Reines-Cowan series of experiments that
showed existence of neutrinos as elementary particles \cite{reines53}.

\begin{figure}[htb]
\centerline{\includegraphics[width=0.6\linewidth]{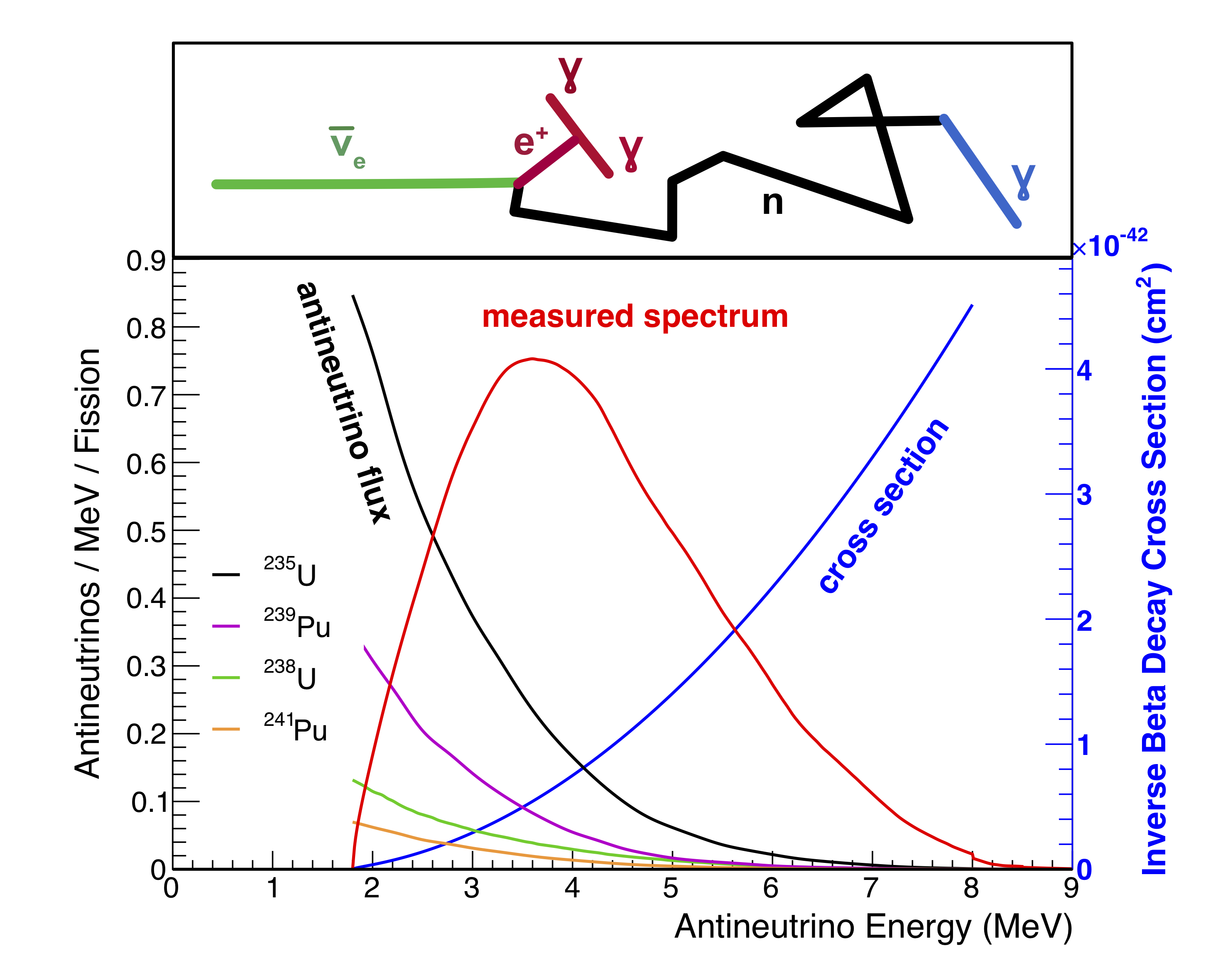}}
\caption{Reactor $\bar{\nu}_e$ flux, inverse-beta-decay cross section, and
$\bar{\nu}_e$ interaction spectrum.
 The flux for individual fuel isotopes, weighed by their typical contribution
 to the total flux in commercial reactors, is also shown. The steps involved in the
 detection are schematically indicated at the top of the figure. Reproduced from \protect\cite{nature15}. }
\label{fig:ibd}
\end{figure}

In fig. \ref{fig:ibd}, reproduced from \cite{nature15}, a typical $\bar{\nu}_e$ interaction spectrum
is shown, the product of decreasing $\bar{\nu}_e$ reactor
spectrum and increasing interaction cross section. The IBD cross section is simply related (in the zeroth approximation)
to the rate of the neutron beta decay $n \rightarrow p + e^- + \bar{\nu}_e$, namely
$\sigma = 2\pi^2/m_e^5 \times E_e p_e /f \tau_n$, where $E_e, p_e$ are the positron energy and momentum,
$f$ is the neutron beta decay phase space integral and $\tau_n$ is the neutron lifetime. Corrections to this
formula, including the effects of the neutron recoil, weak magnetism, and QED are derived, e.g. in ref. \cite{v-b99}.

\section{Exploring the possibilities in the 60s and 70s} 

In the 60s and 70s study of neutrinos was not at the forefront of particle physics. Given the obvious difficulty
of observing neutrino induced reactions, it took the dedication and pioneering spirit of Frederic Reines and his collaborators
to observe all of the above listed reactions with reactor  $\bar{\nu}_e$. The ccd and ncd reactions on deuteron targets
were observed in   \cite{pasierb79} and the $\bar{\nu}_e$ - electron scattering, with the destructive interference
of the charged and neutral currents, and possible electromagnetic interaction if neutrinos
have magnetic moments, were observed in \cite{reines76}. While the measurement of reactions on deuterons 
was never so far repeated by anybody else, the electron scattering is one of the most sensitive probes of the possible
neutrino magnetic moment and is a subject of repeated searches. 

These experiments, with detectors very close to the reactor core, demonstrated that the detection of reactor neutrinos
is possible with detectors on the surface, essentially unprotected from cosmic rays. They also showed that it is possible
to overcome the reactor associated backgrounds. 

\section{Looking for oscillations under the streetlamp in the 80s and 90s}

In late 70s the idea of neutrino oscillations became
a hot subject widely discussed in the community
(note that the famous Physics Reports review by
Bilenky and Pontecorvo appeared in 1978 \cite{bilenky78}.)
As far as the reactor neutrinos are concerned, the only
channel open to experimental study is the $\bar{\nu}_e$
disappearance; the energy is too low that only positrons can be produced, not $\mu^+$ or 
$\tau^+$ leptons. Assuming that the neutrino oscillations
can be described as involving just two neutrino flavors,
the probability that a $\bar{\nu}_e$ of energy $E_{\nu}$ will
change into another flavor neutrino $\bar{\nu}_x$ is
\begin{equation}
P(\bar{\nu}_e \rightarrow  \bar{\nu}_x) = \sin^2 2\theta
\sin^2 \frac{1.27 \Delta m^2 {\rm (eV)}^2 L { \rm (m)}}{E_{\nu} {\rm (MeV)}}~.
\end{equation}
The corresponding phenomenological parameters, the mixing angle $\theta$
and the mass squared difference $\Delta m^2$ were totally unknown
at that time. Hence, experiments were set up `under the streetlamp',
that it where it was relatively easier to perform them.

Reactor experiments with $\sim$ MeV neutrinos were 
reasonably realistic at that time, with  few hundred kg
of the hydrogen containing  detectors
at $L <$ 100 m from the reactor core, that would be
sensitive to the oscillations with $\Delta m^2$ near 1 eV$^2$.
Large number (more that 20) of such experiments
were performed, at that time constraining the $\Delta m^2$ in the indicate
range and a relatively large mixing angles. 

These were all `one detector' experiments, in which the signal
at $L$ was compared to expected signal at $L = 0$, i.e. to the
 spectrum of the $\bar{\nu}_e$ emitted by the reactor. The knowledge
 of that spectrum and its uncertainties was, therefore, an essestial
 ingredient.
 
 I will describe now two of the historically first reactor neutrino oscillation searches.
 
\begin{figure}[htb]
\centerline{\includegraphics[width=0.4\linewidth]{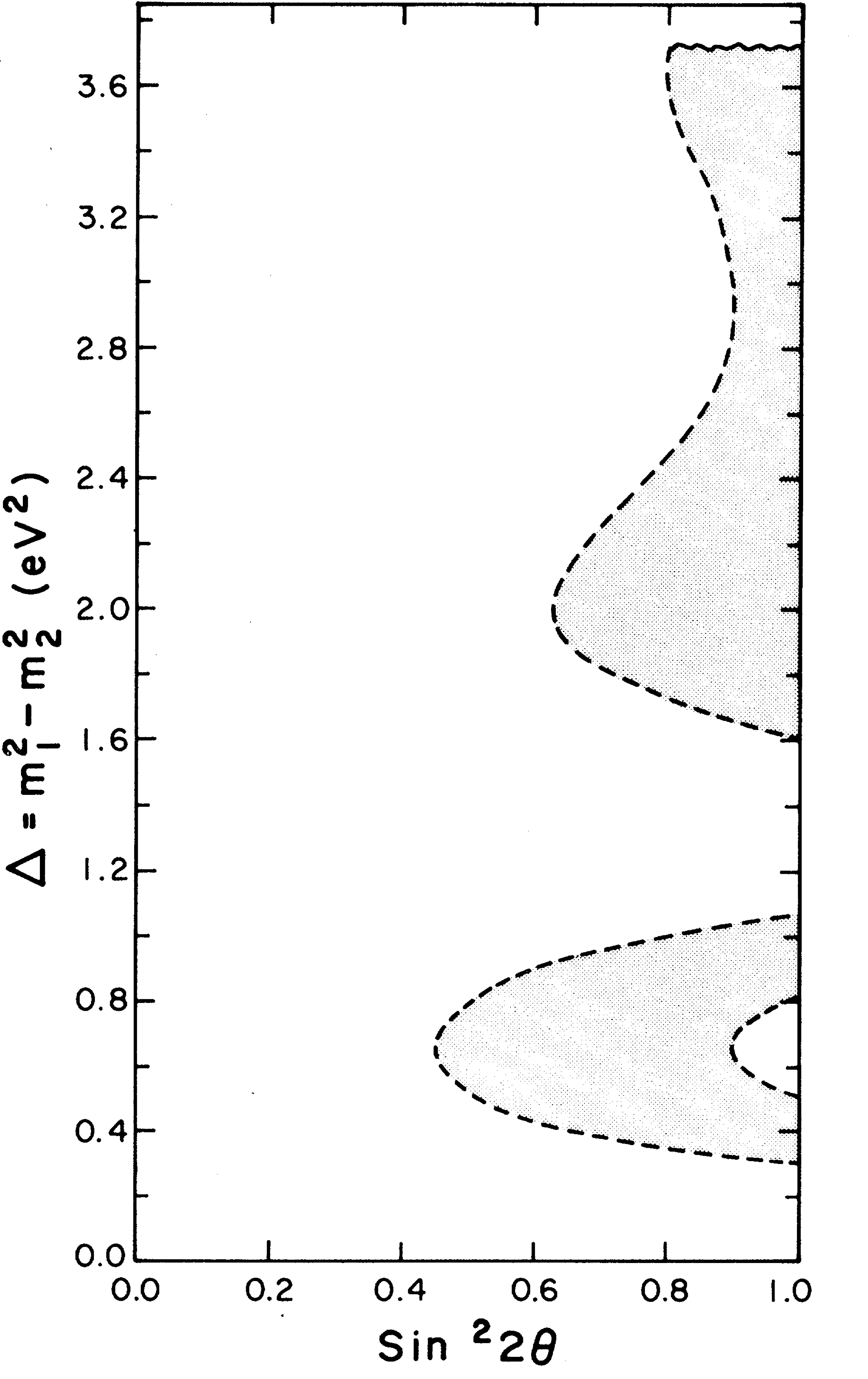}}
\caption{ Grey areas, enclosed in the dashed lines, are the allowed regions of the
parameter space. Reproduced from \protect\cite{reines80}.}
\label{fig:rei}
\end{figure}

 In Ref. \cite{reines80} the rate of the charged and neutral current deuteron
 experiments, $\bar{\nu}_e + d \rightarrow n + n + e^+$ and
 $\bar{\nu}_e + d \rightarrow n + \bar{\nu_e}$ were compared. The
 corresponding thresholds and cross sections are listed in Table I.
 The 268 kg D$_2$O detector was placed 11.2 m from the 2000 MW
 power reactor. Only the neutrons were detected; two neutron events
 represented the charged current and one neutron events represented
 the neutral current. Using the calculated reactor neutrino spectrum \cite{davis79},
 newly evaluated at that time, the rate of the charged current events was
 smaller than expectation, 0.44$\pm$0.19, while the rates of the neutral
 and ccp reactions agreed within errors. This was interpreted as 
 `Evidence for Neutrino Instability'. The corresponding allowed
 region in the $\Delta m^2$ and $\sin^2 2 \theta$ is shown in Fig. \ref{fig:rei},
 reproduced from \cite{reines80}.
 
The paper created a considerable excitement at that time. However, there were also doubts. 
Why was not disappearance noted in the reaction on protons, which were detected actually
at two distances, 6 and 11.2 m ? Also, given the difference in the thresholds of the charged
and neutral current reactions, there was an obvious strong dependence on the correct reactor
spectrum that changes rapidly between 2 and 4 MeV.  The claim was eventually withdrawn
\cite{reines83}; the neutron detection efficiency used was revised. Note, incidentally, that this was
not a genuine planned oscillation experiment. The apparatus and some of the data came from 
Ref. \cite{pasierb79}.
 
The ILL experiment \cite{kwon81} was the first experiment built 
specifically to test the oscillation scenario.
It used the   57 MW research reactor at the Institute Laue Langevin in Grenoble, France
that uses 93\% $^{235}$U enriched fuel. The detector, contained 377 l of liquid scintillator
in 5 walls of target cells with 4 sets of $^3$He wire chambers that detected neutrons in between them.
The detector was placed 8.76 m from the compact reactor core.

\begin{figure}[htb]
\centerline{\includegraphics[width=0.7\linewidth]{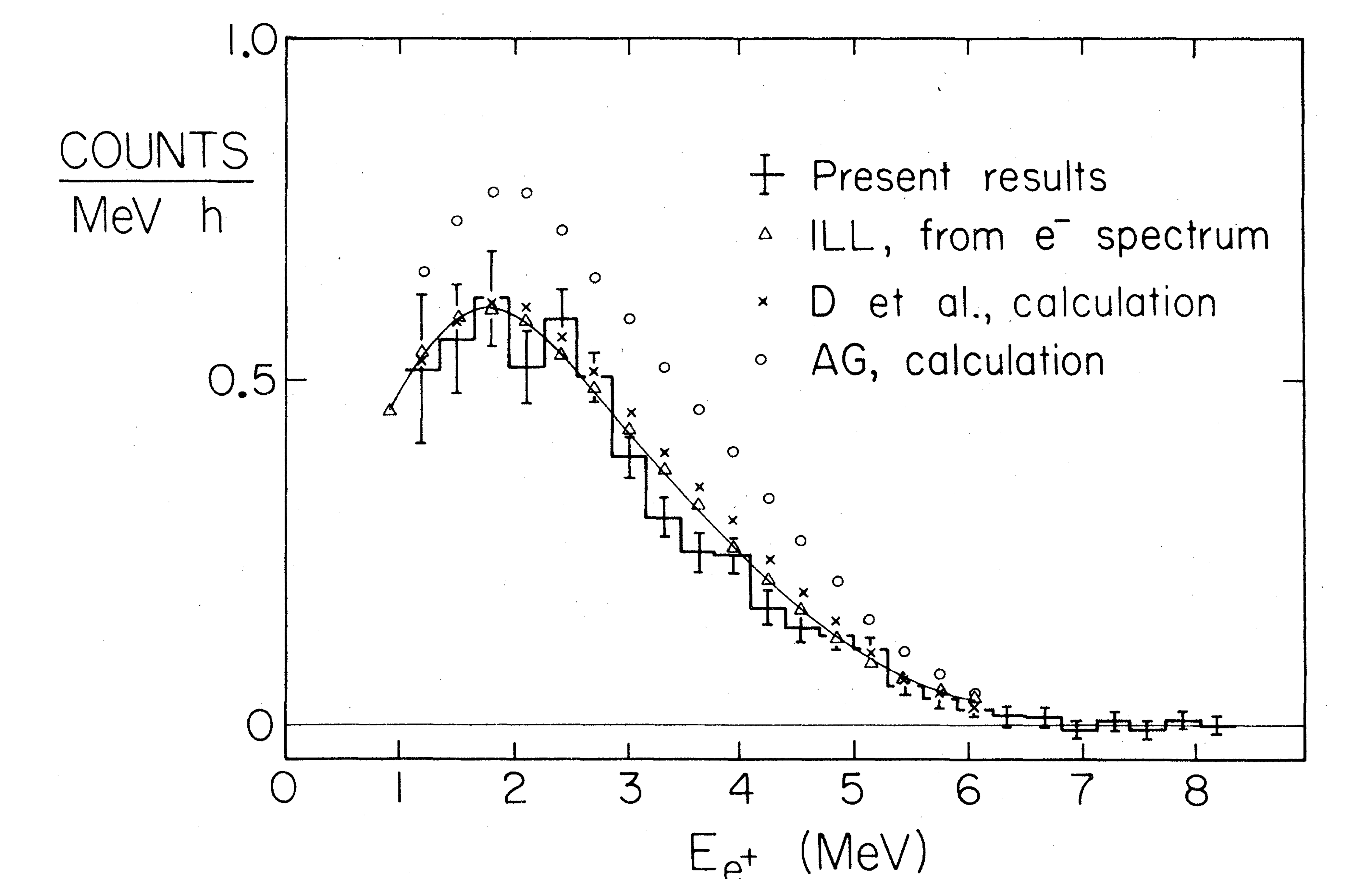}}
\caption{ The positron kinetic energy spectrum. The solid line is based on the $\bar{\nu}_e$ spectrum,
derived from the experimental electron spectrum described later in the text. The D et al. spectrum is
the prediction of ref. \protect\cite{davis79}. The AG spectrum is based on an older evaluation that
obviously predicted too many $\bar{\nu}_e$ above the threshold.  Reproduced from \protect\cite{kwon81}.}
\label{fig:ill}
\end{figure}

The detected positron spectrum, together with expectations for no oscillations, is shown in Fig. \ref{fig:ill}.
It is reproduced from ref. \cite{kwon81}.

 Obviously, no indication of neutrino oscillations was detected. The ratio of the experimental to expected integral
positron yield for $E_{e^+} > 1$ MeV was found to be \\
 $\int Y_{exp}(E_{e^+}) dE_{e+}/ \int Y_{no osc.}(E_{e^+}) dE_{e+}$ = 0.955$\pm$ 0.035(stat) $\pm$ 0.11(syst).
 
However, interestingly that is not the end of the story. In 1995, 15 years later,
 it was announced \cite{houmada95} that the operating power of the high-flux
reactor of Institute Laue Langevin (ILL), Grenoble, had been incorrectly reported
since its earliest days of operation. One impact of this is that the ILL reactor was
operated at the time of the experiment at 1.095 times it's rated full power ( 57 MW thermal).
Therefore, the ILL experiment had to be reanalysed; it showed now a depletion of 17\%
in the neutrino flux. (Neutron lifetime and few other things were changed together
with the reactor power.)
Thus the ratio of experimental to expected integral positron yield is now only
0.832$\pm$ 3.5 \%(stat) $\pm$ 8.87 \%(syst), seemingly indicating $\bar{\nu_e}$
disappearance.  

But this is unlikely so. A number of analogous experiments, described briefly later, does
not show such a substantial depletion. And, in particular, a very similar experiment OSIRIS,
\cite{osiris15} on the reactor with a substantial $^{235}$U enrichment (19.75\%), like at ILL,
 with the detector at 7.21 m from the reactor core, gave $R_{obs}/R_{pred}$ = 1.014 $\pm$ 0.108.
 The ILL result remains unexplained.
 
 Since this conference is about history, it is worthwhile to mention also aspects that are not strictly
 speaking related to physics, but are nevertheless important to the pursuit of physics results. The
 issue is the role of the personal contacts in suggesting, planing, funding, and performing experiments.
 Having good personal contacts considerably helps to some, and not having them is a disadvantage to 
 others. I will illustrate the issue using the ILL experiment as an example. At that time in my home
 institution, Caltech,  there was a very active group of theorists working with Murray Gell-Mann.
 Among them, in particular, Peter Minkowski and Harald Fritzsch were very interested in neutrino
 oscillations and strongly encouraged my colleague Felix Boehm to think about possible experimental
 search. That lead to the idea of a reactor experiment. But where can one find a suitable reactor with
 potentially supportive administration? That lead again to the ILL reactor where Rudolf M\"{o}ssbauer, then professor
 in Munich, Germany, was a current Institute director. He was a former member of Caltech faculty, and
 was not only enthusiasticaly interested, but became, with his students and collaborators an active part of the team 
 of Caltech, Munich and Grenoble physicists who built
 and run the reactor neutrino oscillation experiment \cite{kwon81}. This chain of friendships was essential
 for the success of the experiment.

 \subsection{Reactor spectrum}
 
 Clearly, the knowledge of the reactor neutrino
spectrum is crucial. So, how it could be determined? And how it was determined in the past?
 
 There are two ways, each with its strengths and weaknesses, which were developed during that time pariod: 
 
 1) Add the beta decay spectra of {\bf all} fission fragments.
That obviously requires the knowledge of the contribution of each fuel
 isotope to the reactor power, of the fission
yields (how often is a given isotope produced in fission),
half-lives, branching ratios, and endpoints of all beta branches,
and spectrum shape of each of them. And, naturally also, error bars of
all of that.
 
 Thus, disregarding for clarity the error bars, the neutrino spectrum can be expressed as
 \begin{equation}
 S(E_{\bar{\nu}}) = \frac{W_{therm}}{\Sigma_i (f_i/F) e_i} \Sigma_i \frac{f_i}{F} \left(\frac{dN_i}{dE_{\bar{\nu}}}\right) ~,
 \end{equation}
 where $W_{therm}$ is the thermal power of the reactor, $f_i$ is the number of fissions of the fuel isotope $i$, 
 $F$ is the total number of fissions, $e_i$ is the energy per fission, 
 and $dN_i/dE_{\bar{\nu}}$ is the spectrum of the fuel isotope $i$. In turn
 \begin{equation}
 \frac{dN_i}{dE_{\bar{\nu}}} = \Sigma_n Y_n(Z,A,t) \Sigma_j b_{n,j} P_{\bar{\nu}} (E_{\bar{\nu}} E_0^j, Z) ~,
 \end{equation}
 where $Y_n(Z,A,t)$ is the cumulative yield of the isotope $Z,A$ at time $t$, $b_{n,j}$ is the branching ratio
 of the $\beta$ branch $j$ and $P_{\bar{\nu}} (E_{\bar{\nu}} E_0^j, Z)$ is the normalized spectrum shape of the
 branch $j$ with the endpoint $E_0^j$.
  
 This is seemingly quite straightforward. The thermal power and fission fractions $f_i/F$ are typically supplied by the reactor
 operators and the energy per fission, $e_i$, is known with only small uncertainty. 
 There are libraries of fission yields, but some, in particular for isotopes with short half-lives and hence
 large endpoint energies, have
 large uncertainties. Some $\beta$ branching ratios and endpoint energies are also poorly known, some are even unknown.  
 And the spectrum shapes are well defined for the allowed beta decays (there are, however, substantial uncertainties
 in the next order corrections for the nuclear finite size and weak magnetism that account for a few \% each), 
 but about 25\% of the decays are 
 first forbidden with a more substantial uncertainty in their spectrum. So, the total uncertainty is rather difficult to estimate. 
 
 \begin{figure}[htb]
\centerline{\includegraphics[width=0.6\linewidth]{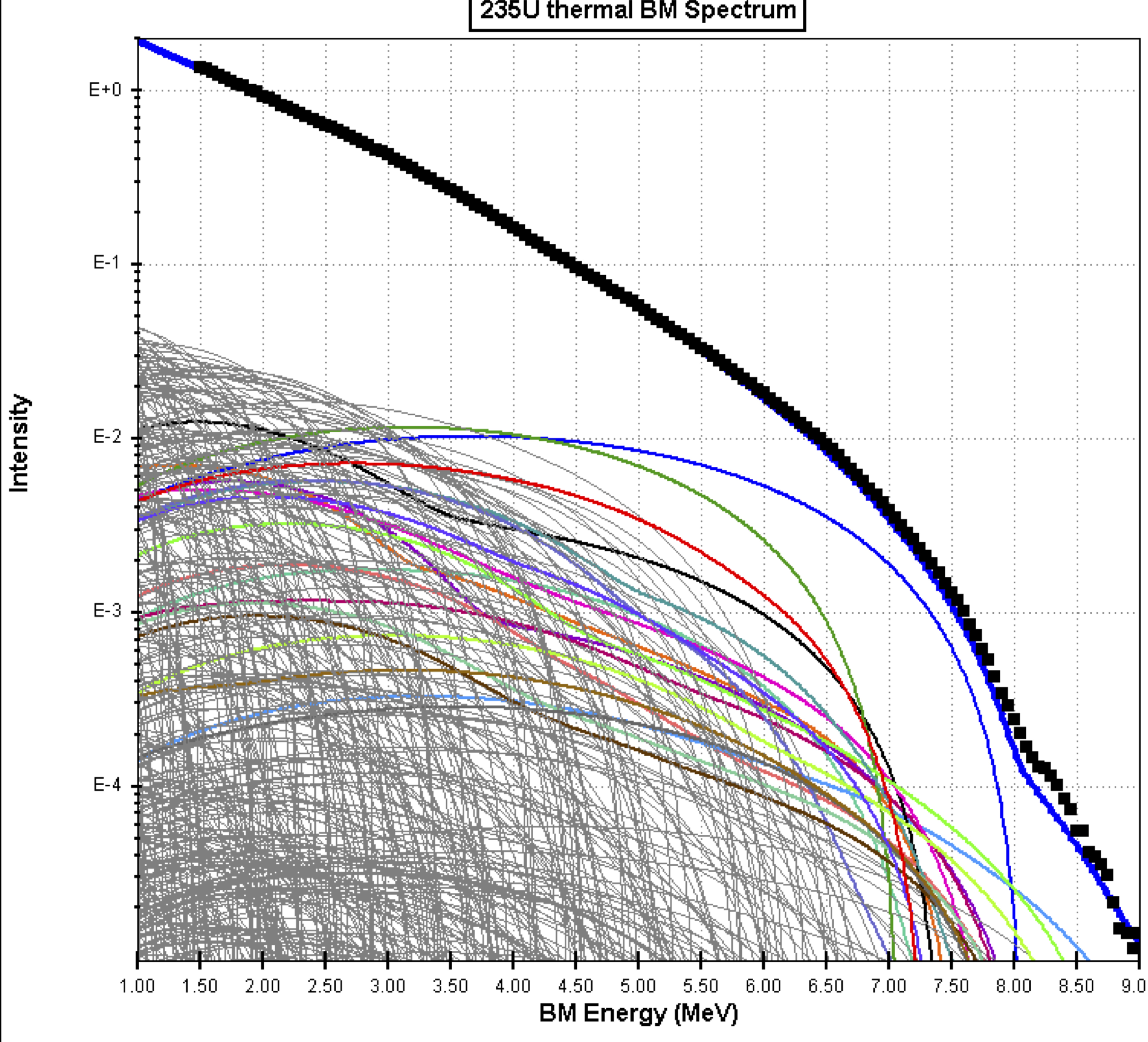}}
\caption{ Calculated electron spectra of the $^{235}$U thermal neutron fission. The thin gray lines
are from individual $\beta$ decays.   The thick (color) lines highlight the 20 most important
contributions to energies above 5.5 MeV. The squares are the sum of all decays and the thick blue line
is the measures electron spectrum.  Reproduced from \protect\cite{sonzogni15}.}
\label{fig:sonzogni}
\end{figure}

How it is done is illustrated in Fig. \ref{fig:sonzogni} reproduced from \cite{sonzogni15}.  There are about 800  
individual fission fragments, with several thousand $\beta$ decay branches displayed. As indicated, at higher
energies, above $\sim$5 MeV, the number of significant fragments is greatly reduced and the relative contribution
of each of them correspondingly enhanced. 

This method was used repeatedly \cite{davis79,vogel81,klapdor82,tengblad89,kopeikin97}, 
by adding newer data and using different ways for treatment the unknown $\beta$ transitions. 
More recently, in ref. \cite{mueller11}, more consistent treatment of corrections to the allowed $\beta$ decay shape,
more careful treatment of the fission fragments that do not achieve equilibrium, use of the recently revised neutron
life-time as well as normalization to the electron spectrum explained next, resulted in the upward  revision
of the total $\bar{\nu}_e$ flux by $\sim$ 6\%, without a substantial change in its shape.
 
2) The second method uses the experimentally determined electron spectrum associated with fission of each of the
reactor fuels. The electron spectrum is then converted into the $\bar{\nu}_e$ spectrum using the fact that in each
$\beta$ decay branch the electron and the $\bar{\nu}_e$ share the available energy $E_0^j$. The electron spectra
associated with the thermal neutron fission of $^{235}$U, $^{239}$Pu and $^{241}$Pu were determined in a series
of experiments \cite{feil82,schreck85,hahn89} at the ILL, Grenoble. $^{238}$U, that contributes $\sim$ 10\%
to the power of a typical commercial reactor, fissions only with fast neutrons. Its electron spectrum was determined
at the reactor in Munich, \cite{haag14} only recently and with substantially larger uncertainties. The conversion from
electrons to $\bar{\nu}_e$ is relatively straightforward, see e.g. \cite{vogel07}, and does not introduce additional
bothersome uncertainty. However, since the fission fragments correspond to a large range of nuclear charges $Z$,
and the $\beta$ spectrum shape depends on $Z$ substantially, it is necessary to use the fission and $\beta$ decay
libraries in order to evaluate the average $\langle Z \rangle$ as a function of the endpoint energy of the fragments.
This information is then used in the conversion procedure.

The $\bar{\nu}_e$ spectrum based on the conversion of the experimental ILL electron spectra were a standard
used for the analysis of reactor oscillation experiments until $\sim$ 2011. At that time, the results of \cite{mueller11},
mentioned earlier, and a new conversion of the ILL spectra in ref. \cite{huber11} appeared. Both of them
\cite{mueller11,huber11} rely on
the measure ILL electron spectra, thus naturally, they give essentially identical total rate and only slightly different
slope of the spectrum. This is used as a standard now. It is remarkable, and a bit unusual, that fundamental input
information is based on essentially unique, and so far not repeated experiment. Current reactor experiments,
however, avoid most of the related uncertainty by using two sets of detector, one (so-called monitor) closer
to the reactor complex, and another one at $L$, further away.

\subsection{Reactor anomaly}

\begin{figure}[htb]
\centerline{\includegraphics[width=0.7\linewidth]{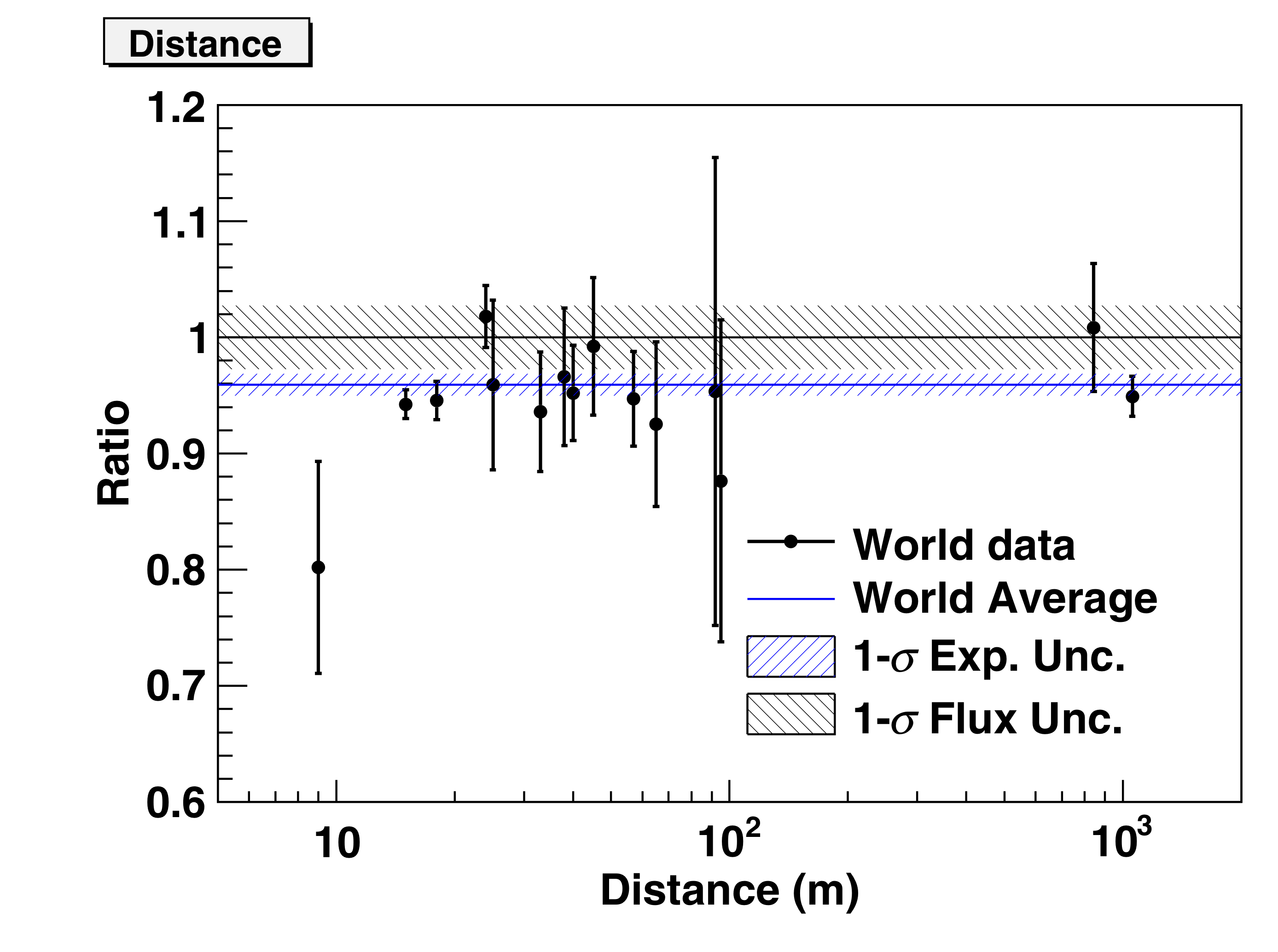}}
\caption{Reactor $\bar{\nu}_e$ capture rate vs. distance $L$ normalized to the
theoretical flux of ref. \protect\cite{mueller11}. The 2.7\% reactor flux uncertainty is 
shown as a band around unity. The (blue) horizontal bar represents the world
average and its $1\sigma$ errors. Data at the same $L$ are averaged for clarity.
Reproduced from \protect\cite{zhang13}.}
\label{fig:anomaly}
\end{figure}

As mentioned earlier, during the 80th and 90th a large number of reactor experiments was performed
at distances $L <$ 100 m
(for the complete list and corresponding references see, e.g. \cite{zhang13}). All of these measurements
are consistent with each other. Resulting total flux is determined to about 1\% experimental accuracy. 
The situation is depicted in Fig. \ref{fig:anomaly}, reproduced from \cite{zhang13} which contains
also results of some newer experiments at $L \sim$ 1 km.

However, the experimentally determined total
flux is $\sim$ 6\% less than the prediction based on \cite{mueller11,huber11} that has an estimated uncertainty
of 2.7\%. This discrepancy became known as the ``reactor anomaly". It can be interpreted as a sign of `new
physics beyond the Standard Model' \cite{mention11}. However, it could also, alternatively,
 mean that the predicted reactor flux is more
uncertain that the estimates in  \cite{mueller11,huber11} suggest, primarily due to the difficulty in accounting
for the shape of the first forbidden $\beta$ decays. There is a lively discussion in the physics community
about this issue and a large number of experiments, some already running at the present time, is designed
to test the beyond the Standard Model hypothesis (see also the contribution of T. Laserre in these proceedings).

\subsection{The ``bump"}

Another unexpected feature in the reactor spectra has been observed in recent reactor experiments
(see the discussion in \cite{hayes16} and the list of original references). The shoulder, centered at $\sim$
5 MeV of the prompt energy, and looking like a broad peak when plotted as the ratio of the observed vs. expected
spectrum, became known as the ``5 MeV bump". It does not affect noticeably the corresponding fit to oscillation parameters,
but casts some additional doubts on the reliability of the calculated reactor $\bar{\nu}_e$ flux.  All experiments that
observe the bump so far employ large volumes of Gd - loaded liquid scintillator,
where the positron energy is recorded together with the two 511 - keV annihilation gamma rays. Is it possible that
the `bump' has something to do with this? Such possibility might explain why an earlier experiment \cite{bugey96}
with a segmented detector, where only the positron kinetic energy is observed, and a different neutron capture 
detection process is used, did not observe such spectrum distortion. 

\begin{figure}[htb]
\centerline{\includegraphics[width=0.55\linewidth]{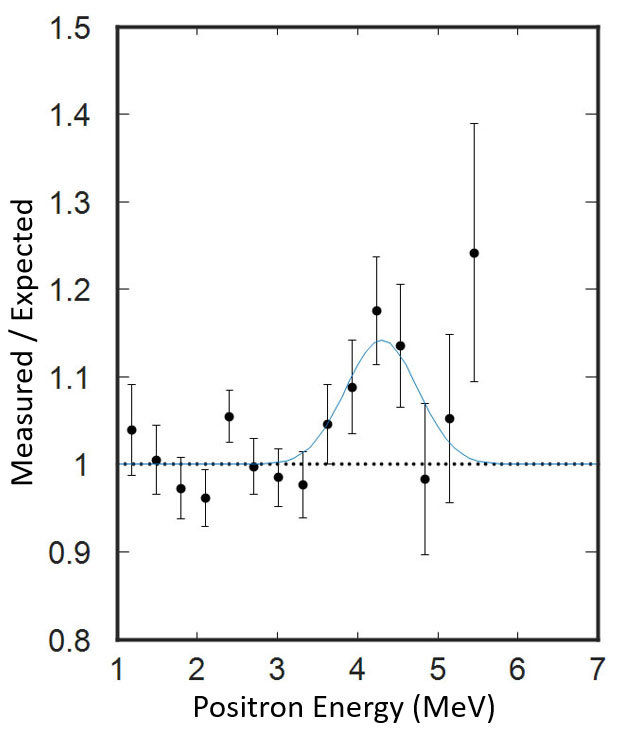}}
\caption{Ratio of the combined measured energy spectrum
(black dots) to the predicted spectrum. The
gaussian fit to the data is indicated. Reproduced from \protect\cite{zacek18}.}
\label{fig:bump}
\end{figure}

However, the reanalysis of the G\"{o}sgen experiment \cite{zacek86}, another experiment with a segmented detector,
shows that the `bump' was in fact present there, as shown in Fig. \ref{fig:bump}, but unfortunately overlooked at that time. 
Moreover, the peak in the ratio plot is now near 4.2 MeV, precisely where it should be if it is indeed caused by the IBD
reaction. Thus, all that points out that indeed the calculated spectra, related all to the ILL 
experiments \cite{feil82,schreck85,hahn89}, somehow miss this feature. 

\section{ Exploring the neutrino oscillations with (almost) known $\Delta m^2$}

In the late 90th the era of blind exploration came to an end. 
The phenomenon of neutrino oscillations was, at least tentatively, established.
Study of atmospheric neutrinos led to the assignment of $\Delta m^2_{atm} \sim (2-4) \times 10^{-3}$
eV$^2$ while the
study of solar neutrinos had, still, several possible solutions, but increasingly
the Large Mixing Angle (LMA) with $\Delta m^2_{sol} \sim 10^{-4}$ eV$^2$
became the preferred one. Both of these disappearance channels were characterized
by relatively large mixing angles thus opening the way for the reactor experiments 
exploring them.

For the reactor neutrino physics it suggested two areas: \\
a) Perform experiments at $\sim$ 1 km corresponding to the $\Delta m^2_{atm}$
and try to determine or constrain the mixing angle $\theta_{13}$. Note that atmospheric neutrinos,
with nearly maximal $\nu_{\mu} \rightarrow \nu_{\tau}$ oscillation probability are insensitive to 
$\theta_{13}$. This program was realized by the Chooz and Palo Verde experiments in the
late 90th and early 2000. This effort and its successful continuation is discussed in detail
by T. Laserre in these proceedings.\\
b) Perform an experiment at $\sim$ 100 km corresponding to the $\Delta m^2_{sol}$
and try to demonstrate the validity of the oscillation interpretation of the
solar neutrino observations also for $\bar{\nu}_e$ at a terrestrial experiment
and without the matter effects.
This program was realized by the KamLAND experiment.

Going from $\sim$ 100 m from the reactor core to $\sim$ 100 km represents an obvious challenge
since the $\bar{\nu}_e$ flux is reduced by the factor of million. The detector must be much larger
and much better shielded against cosmic rays. And all other backgrounds must be substantially reduced
as well. And, in addition, as many reactors as possible should contribute in unison.

\begin{figure}[htb]
\centerline{\includegraphics[width=0.8\linewidth]{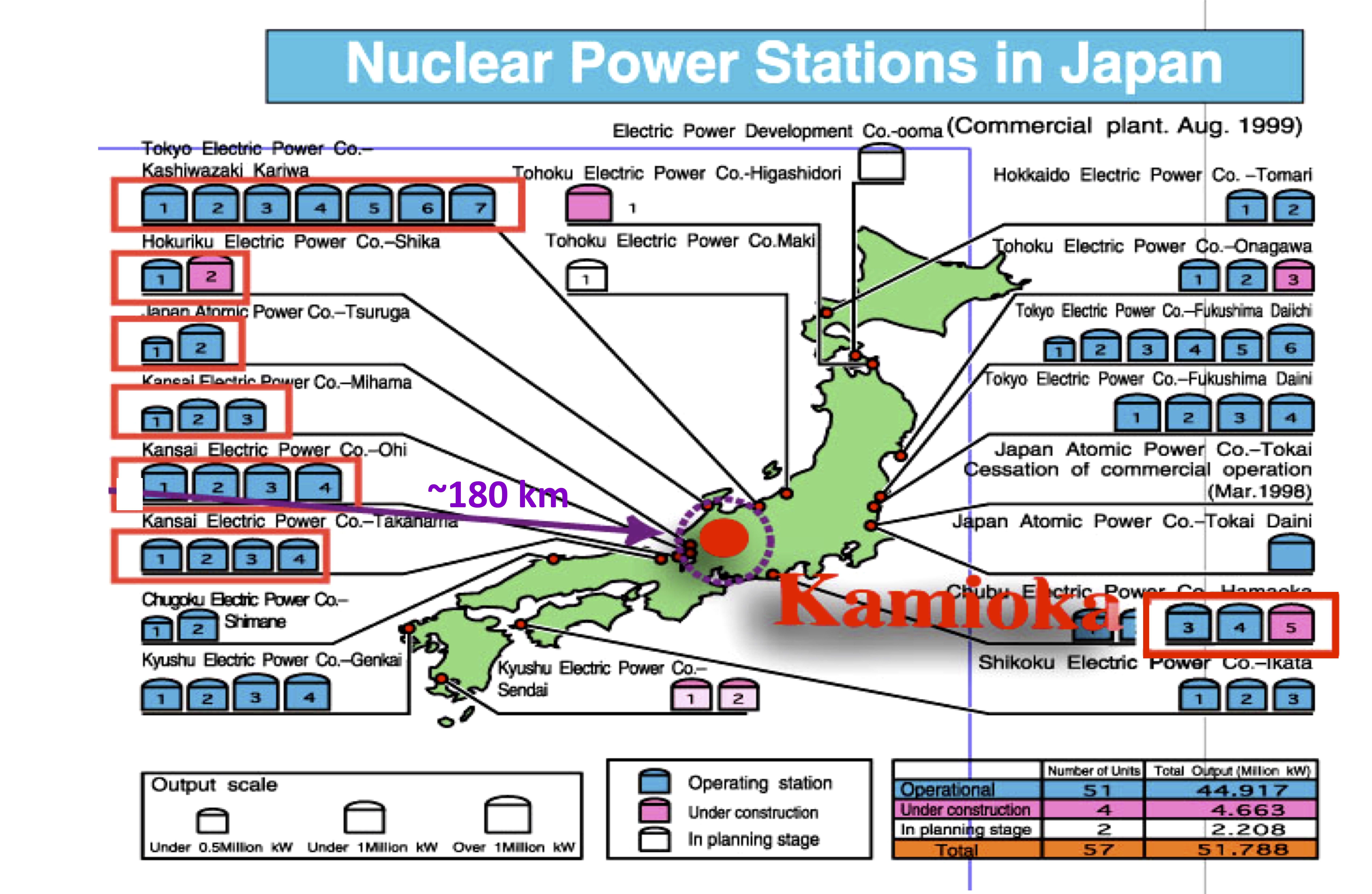}}
\caption{ Nuclear power reactors in Japan. The `magic circle' of reactors around Kamioka
is indicated.}
\label{fig:japan}
\end{figure}

The unique combination of all of that was realized in the KamLAND experiment. At that time
there were $\sim$ 40 working power reactors in Japan. And, by a happy coincidence, many of them were
situated approximately at a circle with radius $\sim$ 180 km centered around Kamioka, as
illustrated in Fig. \ref{fig:japan}. And in Kamioka an underground site became available when
the original Kamiokande experiment was decommissioned.   

KamLAND detector/target  contains 1 kt of ultrapure liquid scintillator; both the $e^+$ and the 2.2 MeV
photons from the neutron capture on hydrogen are detected in the space and time coincidence.
Observing the IBD events was possible thanks to the unprecendented radiopurity of the scintillator liquid.
The amount of $^{238}$U
and $^{232}$Th was reduced to $(3.8 \pm 0.5) \times 10^{-18}$ g/g and $(5.2 \pm 0.8) \times 10^{-17}$ g/g,
respectively. The observed IBD rate was $\sim$ 0.3/ton$\cdot$year, well above the really low background.

\begin{figure}[htb]
\centerline{\includegraphics[width=0.6\linewidth]{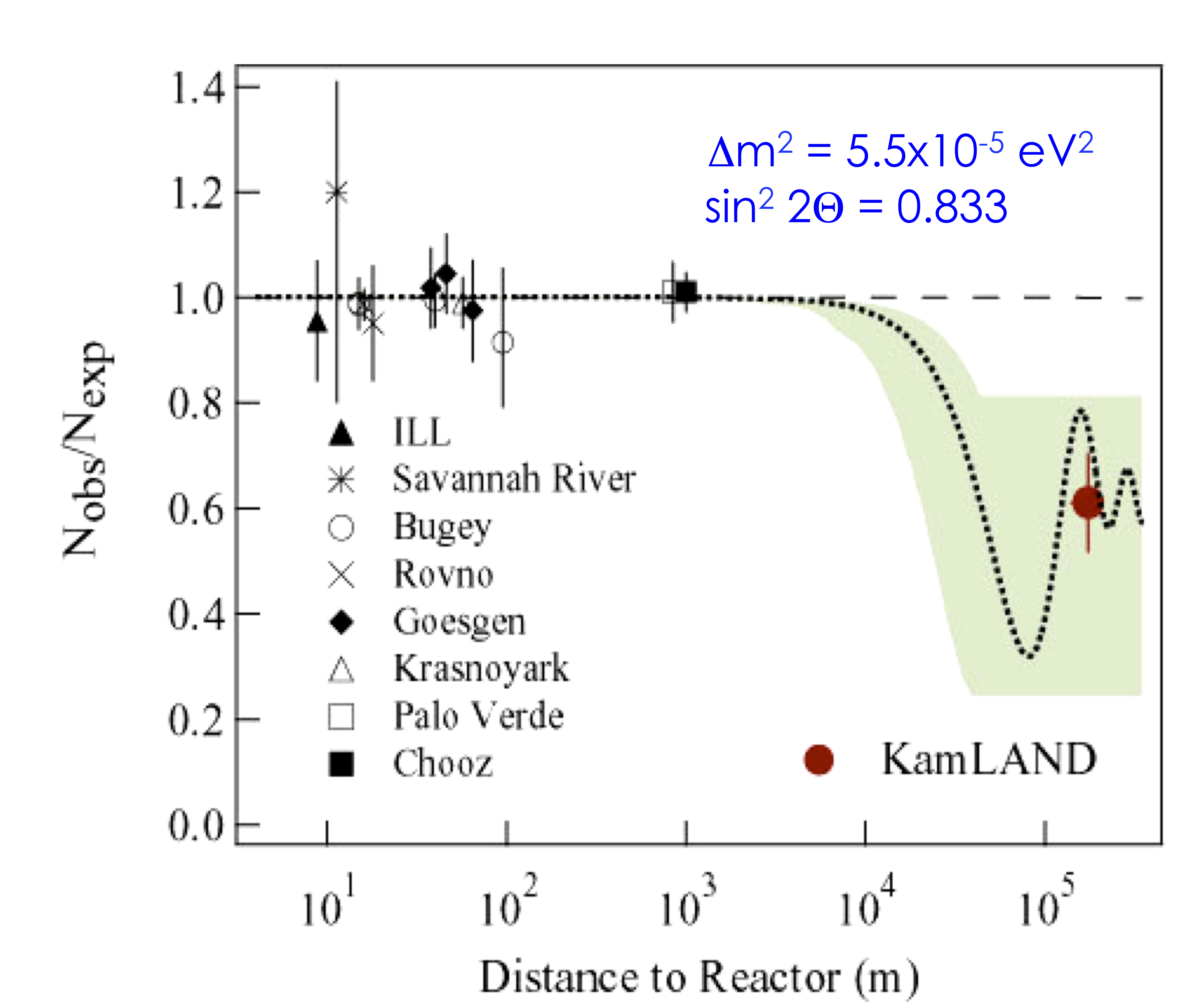}}
\caption{The ratio of measured to expected $\bar{\nu}_e$ flux
from indicated reactor experiments. The solid circle is the
KamLAND result plotted at a flux-weighted average distance
of $\sim$180 km. The shaded region indicates the range of flux
predictions corresponding to the 95\% C.L. LMA region from a
global analysis of the solar neutrino data. The dotted
curve is representative of a best-fit LMA prediction.
Reproduced from \protect\cite{kam1}.}
\label{fig:kam1}
\end{figure}

Already the first KamLAND result \cite{kam1}, using the exposure of 162 ton$\cdot$year, shows clear $\bar{\nu}_e$
disappearance. The ratio of the observed IBD events to the expectations is $0.611 \pm 0.085{\rm ~(stat)} \pm 0.041$
(syst) for $\bar{\nu}_e$ energies above 3.4 MeV. In Fig. \ref{fig:kam1}, reproduced from \cite{kam1}, these
results are plotted together with the results of the previous, shorter baseline, reactor experiments. Not only
is the $\bar{\nu}_e$ disappearance obvious, but it  also demonstrates that this result is in clear agreement with 
the solar neutrino observations, confirming the LMA solutions. 

\begin{figure}[htb]
\centerline{\includegraphics[width=0.8\linewidth]{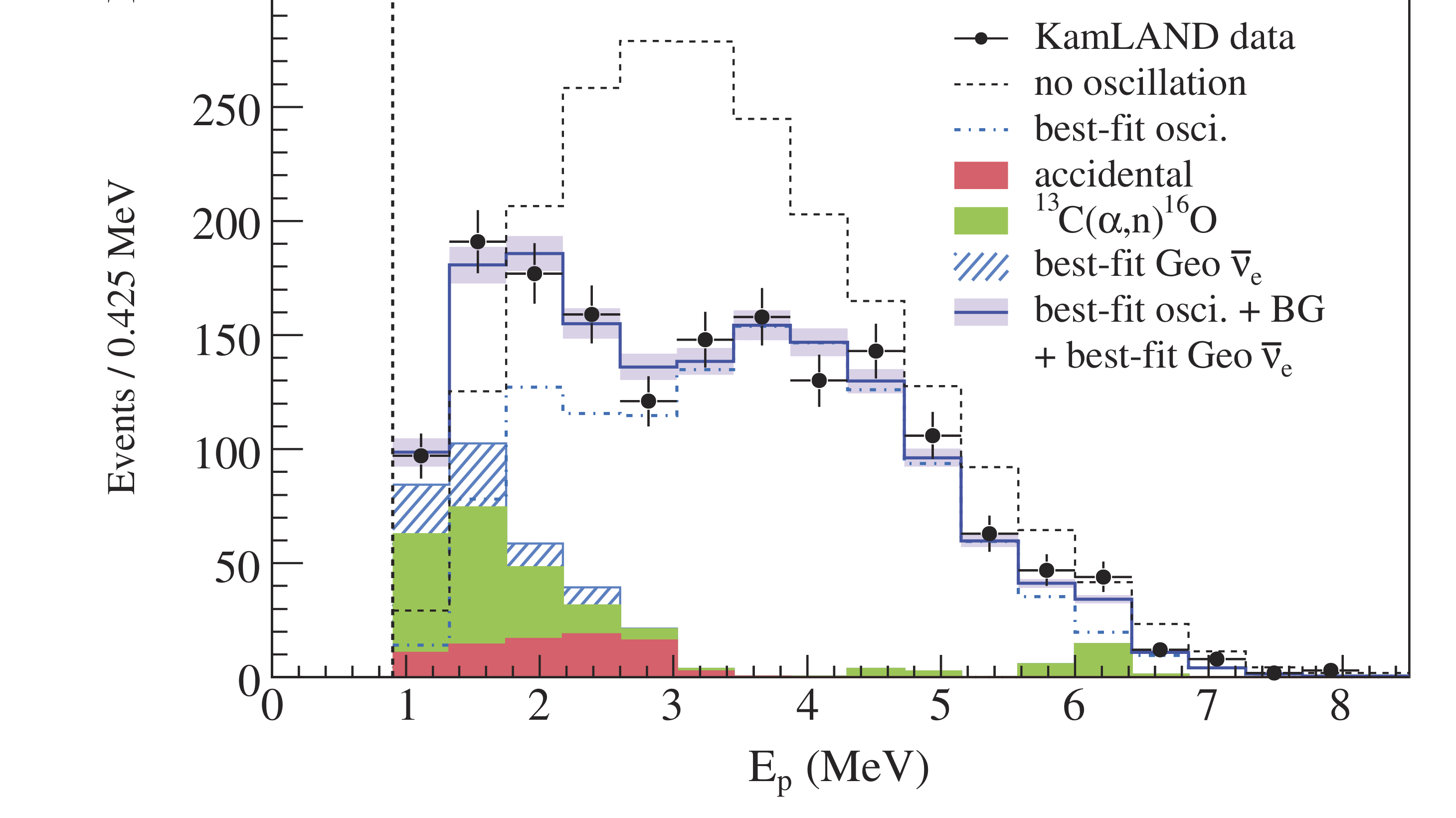}}
\caption{Prompt $\bar{\nu}_e$ energy spectrum. Various background components 
are indicated. The blue histogram indicates the event rate systematic uncertainty.
Reproduced from \protect\cite{kam4}.}
\label{fig:kam4}
\end{figure}

Further exposure with significantly more data (2881 ton$\cdot$year) made it possible to
determine the oscillation parameters with substantially more precision \cite{kam4}. An undistorted
(i.e. no oscillation) reactor $\bar{\nu}_e$ spectrum is now rejected at $> 5 \sigma$. The best fit to the
spectrum gives $\Delta m_{21}^2 = 7.68^{+0.15}_{-0.14} {\rm (stat)} \pm 0.15 {\rm (syst)} \times 10^{-5}$
eV$^2$, i.e. determination with $\sim$ 2.5\% accuracy. Other local minima of $\Delta m_{21}^2$ are
disfavored at $>4 \sigma$, that is the LMA solution is firmly established. 
The experimental reactor $\bar{\nu}_e$ spectrum is compared to the
no oscillation expectation in Fig. \ref{fig:kam4} reproduced from \cite{kam4}. The figure also shows
various background components; the background is well understood.

From the historical perspective another first for KamLAND should be mentioned, even though it
is unrelated to nuclear reactors. For the first
time the so called geoneutrinos,  $\bar{\nu}_e$ produced by the decay of $^{238}$U and $^{232}$Th
within Earth, were detected \cite{geo05}. Earth composition models suggest that the radiogenic power
from these isotopes accounts for approximately half of the total measured heat dissipation rate
from the Earth. Although KamLAND data
have limited statistical power, they nevertheless provide, by direct means, an upper limit (60TW) for the radiogenic
power of U and Th in the Earth, a quantity that is currently poorly constrained.

\begin{figure}[htb]
\centerline{\includegraphics[width=0.8\linewidth]{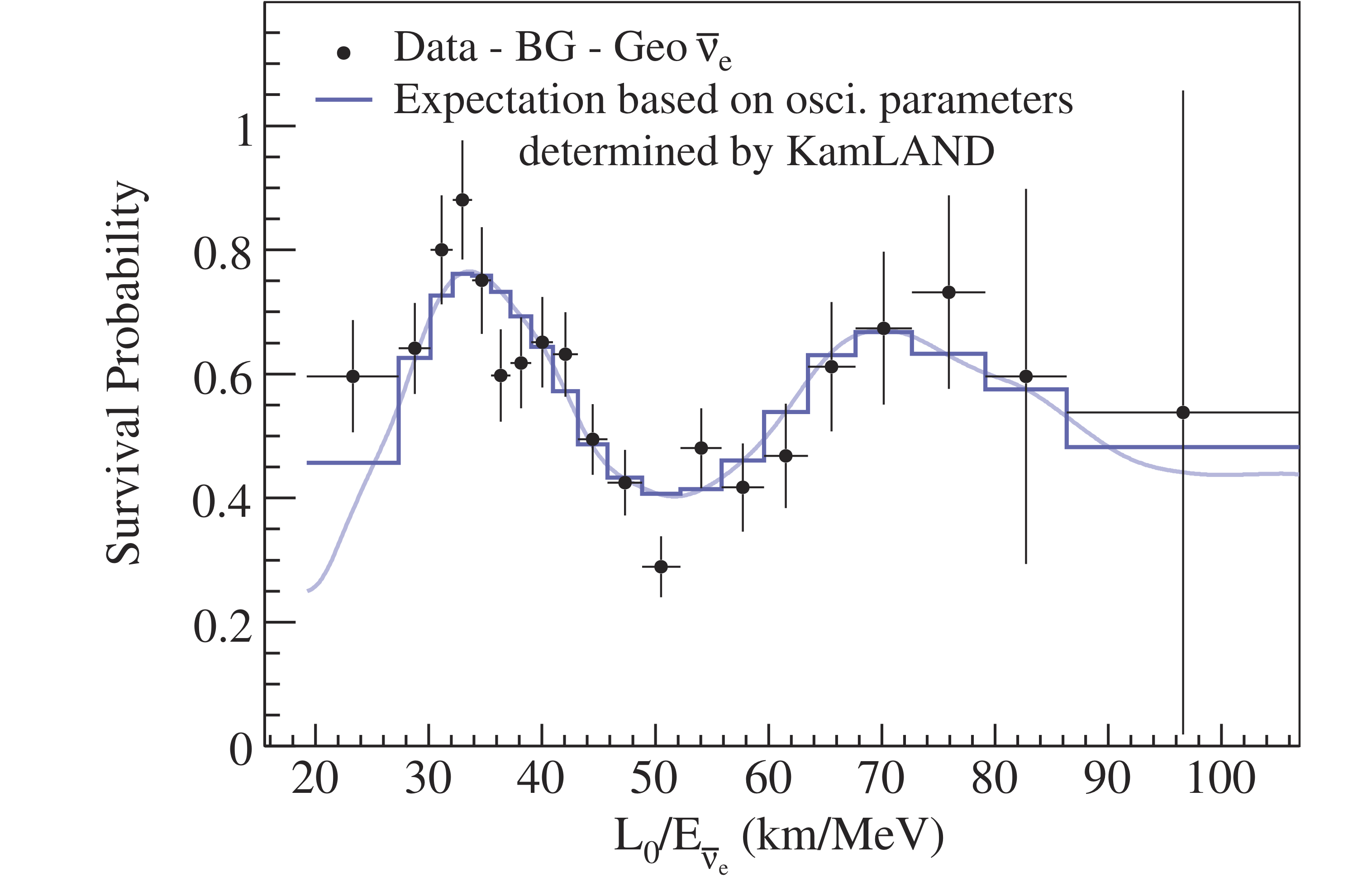}} 
\caption{Ratio of the background and geoneutrino- subtracted $\bar{\nu}_e$ spectrum to the 
expectation for no-oscillation as a function of $L_0/E$. $L_0$ is the effective baseline 
taken as a flux- weighted average ($L_0$ = 180 km). Reproduced from
\protect\cite{kam4}.}
\label{fig:osc}
\end{figure}

The ultimate test of the neutrino oscillation phenomenon is an observation of the periodic changes
of the detection probability of a given neutrino flavor, as prescribed by the eq. (1). Just an observation
of the disappearance of the initial neutrino flavor, or appearance of another one, is a necessary
condition of the oscillation scenario, but not necessarily sufficient one.  As shown in Fig.\ref{fig:osc},
reproduced from \cite{kam4}, KamLAND results show that the disappearance probability of the
reactor $\bar{\nu}_e$ indeed changes periodically as the function of $L/E_{\nu}$. Two full
periods are clearly visible. This was the first time this was accomplished.

There are, naturally, some caveats. The background needs to be subtracted. And the rate has to be
normalized to the no oscillation expectation. And in KamLAND the distance $L$ is not well defined,
since the signal comes from many reactors. However, 86\% of all $\bar{\nu}_e$ events come from the
distance 175$\pm$ 35 km. In fact, the histogram and curve show the expectation accounting for the 
distances to the individual reactors. Thus, the periodic behavior is well supported.

\section{Conclusions}

In this brief contribution I tried to highlight the development of the study of reactor neutrinos over several decades.
Many things happened from the early days  exploring the exotic, incredibly weakly interacting particles, to the present
tool of precision experiments.   And the story is far from over. 
Several large long baseline experiments are being prepared for data taking, with the goal of determining the 
neutrino mass hierarchy, a really challenging task. And the `reactor anomaly' is a hot subject these days,
with many short baseline reactor experiments competing for the goal of either proving that additional sterile neutrinos
indeed exist or that, at least in the context of reactor neutrinos, this is a false alarm. So stay tuned.


\newpage

\section*{References}

\begin{thebibliography}{99}

\bibitem{bethe34} H. Bethe and R. Peierls, Nature {\bf 133}, 532 (1934).

\bibitem{reines53} F. Reines and C. L. Cowan Jr., Phys. Rev. {\bf 92}, 830 (1953);
C.L. Cowan Jr. {\it et al.} Science {\bf 124}, (3212), 103, (1956); F. Reines {\it et al.} 
Phys. Rev. {\bf 117}, 159 (1960).


\bibitem{nature15} P. Vogel, L.J. Wen, C. Zhang, Nature Communications, {\bf 6}, 6935 (2015).

\bibitem{v-b99} P. Vogel and J. F. Beacom, Phys. Rev. D{\bf 60}, 053003 (1999).

\bibitem{pasierb79} E. Pasierb {\it et al.} , Phys. Rev. Lett. {\bf 43}, 96 (1979).

\bibitem{reines76} F. Reines {\it et al.}, Phys. Rev. Lett. {\bf 37}, 315 (1976).

\bibitem{bilenky78} S. M. Bilenky and B. Pontecorvo, Phys, Rep. {\bf 41}, 225 (1978).

\bibitem{reines80} F. Reines, H.W. Sobel and E. Pasierb,  Phys. Rev. Lett. {\bf 45}, 1307 (1980).

\bibitem{davis79} B. R. Davis, P. Vogel, F. M. Mann and R. E. Schenter, Phys. Rev. C {\bf 19}, 2259 (1979).

\bibitem{reines83} F. Reines, Nucl. Phys. {\bf A396}, 469c (1983).

\bibitem{kwon81} H. Kwon {\it et al.} Phys. Rev. D {\bf 24}, 1097 (1981).

\bibitem{houmada95} A. Houmadda {\it et al.} AppL Radiat. Isot. {\bf 46}, 449 (1995).

\bibitem{osiris15} G. Boireau {\it et al.} Phys.Rev.D{\bf 93}, 112006 (2016).

\bibitem{sonzogni15} A. A. Sonzogni, T.D.Johnson and E.A. McCutchan, Phys. Rev.C{\bf 91}, 011301 (2015).

\bibitem{vogel81} P. Vogel, G. K. Schenter, F. M. Mann, and R. E. Schenter, Phys. Rev. C{\bf 24} 1543 (1981).

\bibitem{klapdor82} H. V. Klapdor and J. Metzinger, Phys. Rev. Lett. {\bf 48}, 527 (1982); Phys. Lett. {\bf B112}, 22 (1982).

\bibitem{tengblad89} O. Tengblad {\it et al.}, Nucl. Phys. {bf A503}, 136 (1989).

\bibitem{kopeikin97} V. I. Kopeikin, L. A. Mikaelyan and V. V. Sinev, Phys. At. Nucl. {\bf 60}, 172 (1997).

\bibitem{mueller11} T. A. Mueller {\it et al.} Phys. Rev. C{\bf 83}, 054615 (2011).

\bibitem{feil82} F. v. Feilitzsch, A. A. Hahn, K. Schreckenbach, Phys. Lett. {\bf B118}, 365 (1982).

\bibitem{schreck85} K. Schreckenbach, G. Colvin, W. Gelletly, and F. v. Feilitizsch, Phys. Lett. {\bf B160}, 325 (1985).

\bibitem{hahn89} A. A. Hahn {\it et al.} Phys. Lett. {\bf B218}, 365 (1989).

\bibitem{haag14} N. Haag {\it et al.} Phys. Rev. Lett {\bf 112},122501 (2014).

\bibitem{vogel07} P. Vogel, Phys. Rev. C{\bf 76}, 025504 (2007).

\bibitem{huber11} P. Huber, Phys. Rev. C{\bf 84}, 024617 (2011); erratum, ibid {\bf 85}, 029901 (2012).

\bibitem{mention11} G. Mention {\it et al.}, Phys. Rev. D{\bf 83}, 073006 (2011).

\bibitem{zhang13} C. Zhang, X. Qian and P. Vogel, Phys. Rev. D{\bf 87}, 073018 (2013).

\bibitem{hayes16} A. Hayes and P. Vogel, Annu. Rev. Nucl. Part. Sci.{\bf 66},219 (2016).

\bibitem{bugey96} B. Achbar {\it et al.} Nucl. Phys. {\bf B374}, 243 (1996).

\bibitem{zacek86} G. Zacek {\it et al.}, Phys. Rev. D{ \bf 34}, 2621 (1986).

\bibitem{zacek18} V. Zacek {\it et al.}, Phys. Rev. C (to be published); arXiv:1807.01810.

\bibitem{kam1} K. Eguchi {\it et al.}, Phys. Rev. Lett. {\bf 90}, 021802 (2003).

\bibitem{kam4} S. Abe {\it et al.}, Phys. Rev. Lett. {\bf 100}, 0221803 (2008).

\bibitem{geo05} T. Araki {\it et al.}, Nature {\bf 436}, 499 (2005).

\end{thebibliography}
\end{document}